\documentclass[12pt,draftclsnofoot,onecolumn]{IEEEtran}
\usepackage{graphicx}
\usepackage{fancyhdr}
\usepackage{amsmath,amssymb,amstext}
\usepackage{slashbox}
\usepackage{bm}
\usepackage{algorithm}
\usepackage{array}
\usepackage{cite}

\usepackage{subfigure}
\usepackage{balance}
\usepackage{algorithmicx}
\usepackage{algpseudocode}

\begin{document}

\title{Free-Space Optical Communication Using Non-mode-Selective Photonic Lantern Based Coherent Receiver}

\author{Bo Zhang, Renzhi Yuan\textsuperscript{$\dagger$},~\IEEEmembership{Student Member,~IEEE}, Jianfeng Sun, \\Julian Cheng,~\IEEEmembership{Senior Member,~IEEE}, and Mohamed-Slim Alouini,~\IEEEmembership{Fellow,~IEEE}
\thanks{
Bo Zhang and Jianfeng Sun are with the Key Laboratory of Space Laser Communication and Detection Technology, Shanghai Institute of Optics and Fine Mechanics, Chinese Academy of Sciences, 390 Qinghe Rd., Shanghai 201800, China, (e-mails: zhangbo@siom.ac.cn, sunjianfengs@163.com); Bo Zhang is also with the Center of Materials Science and Optoelectronics Engineering, University of Chinese Academy of Sciences, Beijing 100049, China;
Renzhi Yuan and Julian Cheng are with the School of Engineering, The University of British Columbia, Kelowna, V1V 1V7, BC, Canada (e-mails: renzhi.yuan@ubc.ca, julian.cheng@ubc.ca);
Mohamed-Slim Alouini is with the Computer, Electrical, and Mathematical Science and Engineering (CEMSE) Division King Abdullah University of Science and Technology (KAUST) Thuwal, Makkah Province, Saudi Arabia (e-mail: slim.alouini@kaust.edu.sa).}
\thanks{$\dagger$: Corresponding author}
}

\maketitle

\begin{abstract}
A free-space optical communication system using non-mode-selective photonic lantern (PL) based coherent receiver is studied. Based on the simulation of photon distribution, the power distribution at the single-mode fiber end of the PL is quantitatively described as a truncated Gaussian distribution over a simplex. The signal-to-noise ratios (SNRs) for the communication system using PL based receiver are analyzed using different combining techniques, including selection combining (SC), equal-gain combining (EGC), and maximal-ratio combining (MRC). The integral solution, series lower bound solution and asymptotic solution are presented for bit-error rate (BER) of PL based receiver, single-mode fiber receiver and multimode fiber receiver over the Gamma-Gamma atmosphere turbulence channels. We demonstrate that the power distribution of the PL has no effect on the SNR and BER performance of the PL based receiver when MRC is used; and it only has limited influence when EGC is used. However, the power distribution of the PL can greatly affect the BER performance when SC is used. Besides, the SNR gains of the PL based receiver using EGC over single-mode fiber receiver and multimode fiber receiver are numerically studied under different imperfect device parameters; and the scope of application of the communication system is further provided.
\end{abstract}


\IEEEpeerreviewmaketitle

\section{Introduction}
In satellite communication, coherent free-space optical communication (FSOC) technology is attractive for its high sensitivity and ability to obtain a high data rate \cite{liu2015new,shemis2017self,zhou2017symbol}. Recently, researchers have focused on designing coherent optical communication systems using fiber-based transmitters and receivers. Because the fiber-based receiver can make full use of the commercial available components from fiber-optic communication systems, such as fiber transmitter and receiver, erbium-doped fiber amplifiers (EDFAs), and fiber multiplexer and demultiplexer units \cite{dikmelik2005fiber, zhang2013fiber}. However, such implementation has its limitations because the overall efficiency (which will be defined in the sequel) is low.

In a coherent optical communication system using fiber-based receiver, when the signal beam reaches the receiver aperture plane, it is first coupled into the fiber and then mixed with the local oscillator (LO) beam to obtain the mixed signal. There are three important parameters associated with this process: coupling efficiency, mixing efficiency, and overall efficiency. The coupling efficiency is defined as the ratio of the average power coupled into the fiber to the average power in the receiver's aperture plane \cite{winzer1998fiber,dikmelik2005fiber,toyoshima2006maximum,grein2014multimode,poliak2016fiber,arisa2017coupling,hu2018fiber}. The mixing efficiency is defined as the ratio of the amplitude of the obtained mixed signal to the amplitude of theoretically mixed signal \cite{mark1992comparison,duncan1993performance,jacob1995heterodyne,leeb1998aperture,ren2012heterodyne}. The overall efficiency is defined as the product of the coupling efficiency and the mixing efficiency, which embodies the extent to which the signal beam can be fully utilized. Low overall efficiency can typically degrade signal-to-noise ratio (SNR) \cite{duncan1993performance,jacob1995heterodyne}.

There are two commonly used fiber-based receiver schemes for coherent optical communication systems. The first receiver scheme is the single-mode fiber (SMF) receiver with SMF mixing. The SMF only propagates one field mode. Because the received signal beam and the LO beam propagate in the same SMF, their field modes are the same, i.e., the received signal beam and LO beam fields are matched both spatially and temporally at the detector. Then the mixing efficiency between the LO beam and the signal beam approaches 100\% \cite{jacob1995heterodyne}. However, the core diameter of SMF is small ($\sim 10 \ \mu m$), which limits achievable fiber coupling efficiency, especially in the presence of atmosphere turbulence in free-space channels \cite{mark1992comparison,winzer1998fiber,dikmelik2005fiber,toyoshima2006maximum}. For example, the maximum coupling efficiency is 81\% in the absence of atmosphere turbulence \cite{winzer1998fiber}. For a moderate strength turbulence ($C_n^2=10^{-13}$ m$^{-\frac 2 3}$, where $C_n^2$ is refractive-index structure constant), the coupling efficiency is less than 5\% \cite{dikmelik2005fiber}. The second receiver scheme is the multimode fiber (MMF) receiver with MMF mixing \cite{duncan1993performance, poliak2016fiber, grein2014multimode}. The coupling efficiency of MMF, whose core diameter is $\sim 50 \ \mu m$ \cite{poliak2016fiber}, is much higher than that of the SMF \cite{duncan1993performance,niu2007coupling,takayama2011studies,zheng2016free,arisa2017coupling}. However, only the portion of the signal beam that is in the same temporal and spatial mode of the LO beam can produce high mixing efficiency \cite{duncan1993performance}. The MMF contains not only the fundamental mode component, but also high-order mode components. Then the mixing efficiency between the LO beam and the signal beam will be degraded \cite{duncan1993performance, ozdur2015photonic, zheng2016free}. For example, the coupling efficiency of MMFs tested in \cite{poliak2016fiber} is greater than 95\%; for asymmetric square waveguide supporting seventy-five distinct modes tested in \cite{duncan1993performance}, the coupling efficiency for MMF receiver is 75-78\%; the mixing efficiency is 21-23\%, and the overall efficiency becomes only 11-17\%. The properties of the SMF receiver and MMF receiver are summarized in Table \ref{tab:my_label}. From Table \ref{tab:my_label}, we can conclude that both SMF receiver and MMF receiver have low overall efficiency.

\begin{table}
\centering
\caption{Comparison between SMF receiver and MMF receiver.}
\begin{tabular}{c c c}
\hline
& SMF receiver& MMF receiver\\
\hline
The coupling efficiencies & low & high \\
The mixing efficiencies & high & low \\
The overall efficiencies & low & low \\
\hline
\end{tabular}
\label{tab:my_label}
\end{table}

\begin{figure}
\begin{center}
\includegraphics[width=0.5\textwidth, draft=false]{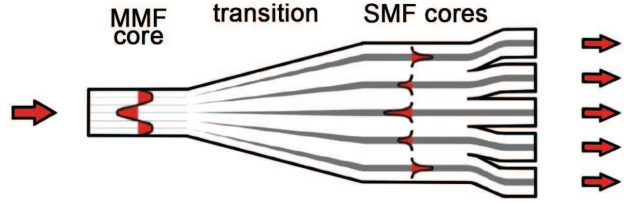}
\caption{Structural diagram of a PL \cite{leon2005multimode}}
\label{fig: PL}
\end{center}
\end{figure}

\begin{figure*}
\begin{center}
\includegraphics[width=0.95\textwidth, draft=false]{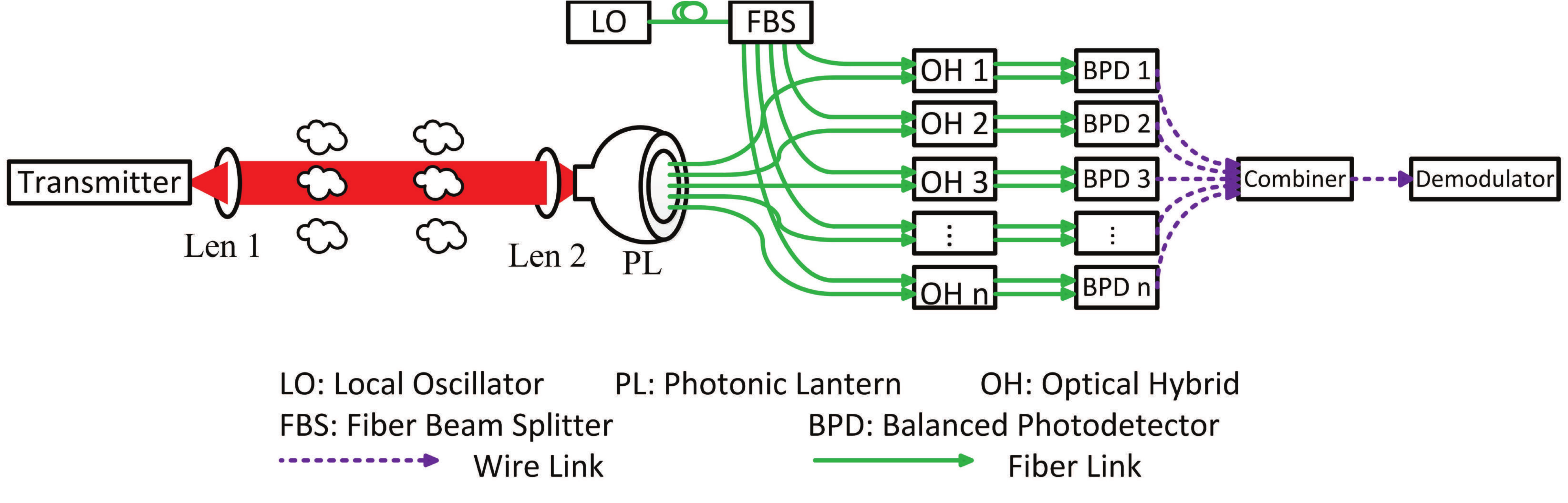}
\caption{Structural diagram of a coherent FSOC system based on non-mode-selective PL}
\label{fig: PL_based_FSOC}
\end{center}
\end{figure*}

Recently, a non-mode-selective photonic lantern (PL) based coherent optical receiver has been proposed, and the overall efficiency of this receiver can be improved \cite{ozdur2013free,ozdur2014performance,ryf2014photonic,ozdur2015photonic,zhang2019study}. Fig. \ref{fig: PL} shows the schematic diagram of a PL \cite{leon2005multimode,birks2015photonic}. In this diagram, one end of the PL is a relatively-large multimode core, and the other end is an array of several relatively-small single-mode cores. In between is a transition region \footnote{There are two types of PLs \cite{huang2015all,birks2015photonic}: mode-selective PL \cite{leon2014mode,li2019mode} and non-mode-selective PL \cite{ozdur2013free,ozdur2014performance,ryf2014photonic,ozdur2015photonic}. In a mode-selective PL, the single-mode cores are designed for transmitting light with different electromagnetic wave modes. While in a non-mode-selective PL, the single-mode cores are designed for transmitting light having the same electromagnetic wave mode. This paper focuses on the non-mode-selective PL.}. Fig. \ref{fig: PL_based_FSOC} shows a structural diagram of a complete coherent FSOC system based on non-mode-selective PL. The signal beam is transmitted from the transmitter, and is coupled into the receiver after passing through the atmosphere turbulence. In the receiver, the large-core MMF end of the PL is placed behind the receiver len to collect the multimode signal beam. Then the PL converts the multimode signal beam into $N$ single-mode signal beams. The single-mode LO beam is split into $N$ equal parts by a fiber beam splitter (FBS). Each single-mode signal beam of the PL is mixed with a single-mode LO beam in an optical hybrid. After that, each mixed signal is converted into an electrical signal by the corresponding balanced photodetector. All the electrical signals are sent to the combiner and the demodulator for processing. The system can fully take advantage of the MMF, which has higher coupling efficiency compared with the SMF, and can take advantage of the SMF that has nearly 100\% mixing efficiency with the single-mode LO beam \cite{ozdur2013free}.

A recent work \cite{zheng2018performance} investigated the performance of coherent FSOC receiver under moderate-to-strong turbulence. However, the effect of the power distribution at SMF end of the PL on SNR was not studied. From \cite{birks2015photonic}, we know that the power distribution at SMF end of non-mode-selective PL varies according to input mode profiles, input pointing errors, and temperature or pressure variations on the MMF section of the PL. In FSOC, the signal beam impaired by atmospheric turbulence contains not only the fundamental mode component but also the higher-order mode components, and the influence of atmospheric turbulence on signal beam changes with time and space. Then the mode profile of the signal beam coupled into MMF end of the PL will change with time, resulting in the power distribution variation at SMF end of a non-mode-selective PL. As a result, the SNR of the coherent optical receiver based on PL will change \cite{zhang2019study}. Therefore, it is necessary to study the power distribution at SMF end of the PL. In \cite{zhang2019study}, we proposed two different distributions: the multivariate Gaussian distribution over a simplex for small power fluctuation case and the uniform distribution over a simplex for large power fluctuation case, to describe the power distribution at SMF end of non-mode selective PL. It was found that different power distributions can have different effects on SNR, and when the number of single-mode fibers of the PL is equal to the number of guided modes at multimode end of the PL, the average SNR attains its maximum value \cite{zhang2019study}.

Different from \cite{zhang2019study}, this paper proposes a more accurate power distribution: truncated Gaussian distribution over a simplex, and this proposal is based on the simulation results of photon distribution in Section \ref{section: Power Distribution }. In addition, the SNRs of the communication system using PL based receiver are analyzed using different combining techniques, including selection combining (SC), equal-gain combining (EGC), and maximal-ratio combining (MRC); and they are compared with traditional SMF and MMF receivers in Section \ref{section: SNR}. The bit-error rate (BER) performance of a binary phase-shift keying (BPSK) system is analyzed. The integral solution, lower bound series solution and asymptotic solution of the BER are presented. The BER performance of the system using different combining techniques are studied and compared with the traditional SMF and MMF receivers over the Gamma-Gamma atmosphere turbulence channels in Section \ref{section: BER Analysis}. We demonstrate that the power distribution of the PL has no effect on the SNR and BER performance of the PL based receiver when MRC is used. Simulation results in Section \ref{section: Simulation} show that the power distribution of the PL has only limited influence on the BER performance of PL based receiver when EGC is used; and the power distribution can greatly affect the BER performance when SC is used. Besides, the SNR gains of the PL based receiver using EGC over the SMF receiver and MMF receivers are numerically calculated under different imperfect device parameters; and the scope of application of the communication system is further provided. To the best of the authors' knowledge, this is the first analytical study on the influences of the power distribution of the PL on the performance of different combining techniques. Our findings can provide some useful guidelines for the design of PL based receiver for FSOC systems.

\section{System Model}
\label{section: SystemModel}
\subsection{Free-Space Atmosphere Channel}
\label{section: Free-space Atmosphere Channel}
In FSOC, atmospheric turbulence introduces fluctuation of irradiance, which results in fluctuation of SNR. The probability density function (PDF) of the received signal irradiance can be modeled as a Gamma-Gamma distribution \cite{al2001mathematical, chatzidiamantis2011distribution}, which emerges as a useful turbulence model as it has excellent fit with measurement data over a wide range of turbulence conditions \cite{al2001mathematical}. The PDF of the received signal irradiance $I \ (I>0)$ is given by
\begin{equation}
f \left ({I} \right )=\frac{2{{ \left (\alpha \beta \right )}^{\frac{\alpha +\beta}{2}}}}{\Gamma \left (\alpha \right )\Gamma \left (\beta \right )}{ {I} ^{\frac{\alpha +\beta}{2}-1}}{{K}_{\alpha -\beta}} \left (2\sqrt{\alpha \beta {{I}}} \right ),
\label{eq: f}
\end{equation}
\noindent where $\Gamma \left (\cdot \right )$ is the Gamma function; ${K_{\alpha -\beta}} \left (\cdot \right )$ is the modified Bessel function of the second kind with order $\alpha -\beta $. The parameters $\alpha $ and $\beta $ are directly related to the atmospheric conditions \cite{al2001mathematical}, and they respectively denote the effective numbers of large-scale and small-scale cells of the scattering process, respectively. Without loss of generality, the received signal irradiance $I$ is normalized, i.e., $E[I]=1$, where $E[\cdot]$ denotes the mathematical expectation.

\subsection{Power Distribution in PL}
\label{section: Power Distribution }
When the signal beam transmitted from the transmitter reaches the receiver system after passing through the atmosphere turbulence, it is coupled into the MMF end of the PL. We assume the power received at MMF end of the PL is ${{P}_{M}}$, then we have \cite{li2016performance}
\begin{equation}
P_M={{\zeta}_{M}} {A} {I},
\label{P_M}
\end{equation}
\noindent where ${\zeta}_{M}$ is the coupling efficiency of MMF, and $A$ is the area of receiving aperture of the len. When the PL converts the multimode signal beam into $N$ single-mode signal beams, loss will be introduced \cite{birks2015photonic}. If we denote the loss factor of the PL by $\xi_{PL} \ (0 < \xi_{PL} \leq 1)$, then the output optical power of the PL is ${{P}_{S}}={{\xi}_{PL}}{{P}_M}$.

For a PL with $N$ SMFs, if we denote the power distributed at each SMF end by ${{P}_{S, i}} \ (i=1, 2, \cdots, N) $ and denote the ratio of ${{P}_{S, i}}$ to ${{P}_S}$ by $a_i$, then we have
\begin{equation}
{P}_{S, i}=a_i P_S = a_i \xi_{PL} \zeta_M A I,
\label{P_S_i}
\end{equation}
\noindent where random variables (RVs) $a_i \ (i=1, 2, \cdots, N)$ satisfy
\begin{equation}
a_1+a_2+\cdots+a_N=1, \quad 0 \leq a_i \leq 1, \quad i=1,2, \cdots, N,
\label{Simplex}
\end{equation}
\noindent where the set of $\{a_1, a_2, \cdots, a_N\}$ that satisfies \eqref{Simplex} is called a standard unit simplex \cite{onn2011generating}.

The exact power distribution at SMF end of the PL is not known. Because the optical power is proportional to the photon number, the ratios $\{a_1, a_2, \cdots, a_N\}$ for the optical power is identical to the ratios for the photon numbers. Therefore, we can simulate the photon distribution to obtain the power distribution.

\subsubsection{Simulation Model For Photon Distribution}
\label{section: Photon Distribution}
Here, we use a Monte-Carlo method to simulate the photon distribution at SMF end of the PL. We denote the number of SMF of the PL by $N$. Because the loss of PL has no effect on the power distribution at SMF end of the PL, we do not consider the loss of PL in the simulation of photon distribution. Because this work assumes non-mode-selective PL, it is reasonable to assume that each SMF of a PL is exactly the same. Then the probability of each photon at MMF end assigned to any SMF of PL is assumed the same. Therefore, the explicit Monte-Carlo process can be summarized as follows: Step 1, we first generate $M$ photons and assign each photon into one SMF end randomly; Step 2, we calculate and record the ratio of the photon number $m_i$ of $i$th SMF end to the total photon number $M$ as $a_i=m_i/M$, where $i=1, 2, \cdots, N$; Step 3, repeat Step 1 and Step 2 $L$ times. Then we can obtain the distribution of $a_i$ from its $L$ samples for the $i$th SMF and obtain the correlation coefficient between $a_i$ and $a_j$ for $i \neq j$.

\begin{figure}
\begin{center}
\includegraphics [width=0.6\textwidth, draft=false] {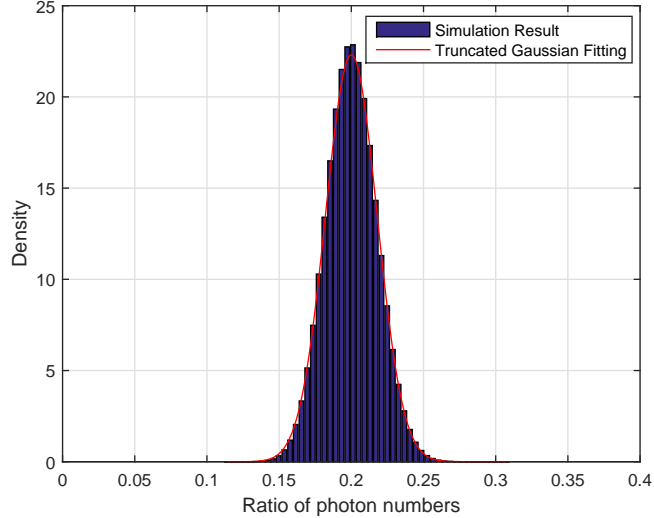}
\caption{The obtained ratio distribution for one SMF end of the PL with $N=5$,  $M=500$, and $L=10^7$ (The range of $a_i$ is between 0 and 1. For simplicity, we only plot the range of $a_i$ from 0 to 0.4.)}
\label{fig: Gau_shape}
\end{center}
\end{figure}

\begin{figure}
\begin{center}
\includegraphics [width=0.6\textwidth, draft=false] {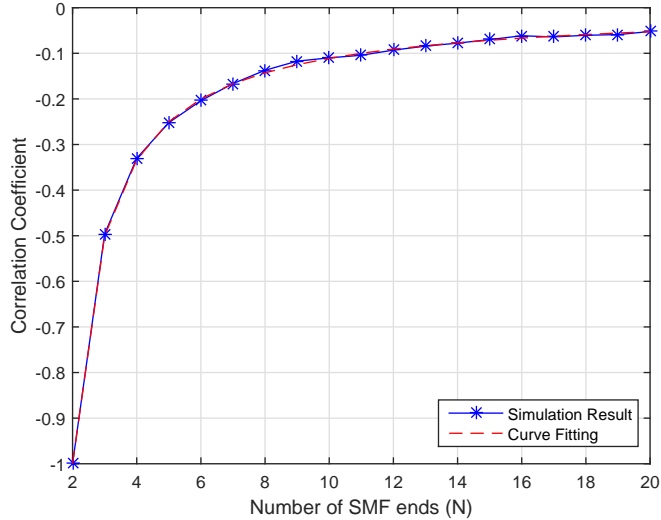}
\caption{The obtained correlation coefficient between the ratios of two distinct SMF with $M=100\times N$ and $L=10^7$}
\label{fig: Correlation_Coefficient}
\end{center}
\end{figure}

The obtained distribution of the ratio $a_i$ for some SMF end is shown in Fig. \ref{fig: Gau_shape}. We find that the photon number distribution at the $i$th SMF end of the PL has excellent fit with the truncated Gaussian distribution with mean value $1/N$ \footnote{We remark that the obtained variance of $a_i$ can vary as the number of simulation repeating times varies due to the converging property of the Monte-Carlo method. A large number of repeating times results in a small variance. However the correlation coefficient between $a_i$ and $a_j$ is independent of the number of repeating times}. The obtained correlation coefficients between the ratios of two distinct SMF over the number of SMF ends are shown in Fig. \ref{fig: Correlation_Coefficient}. We can see that the correlation coefficients between the ratios of two distinct SMFs are always negative, which is due to the constraint \eqref{Simplex}. Besides, we can see that the correlation coefficient between two SMFs increases as $N$ increases. For example, when $N=2$, according to the constraint \eqref{Simplex}, the correlation coefficient between two SMFs is $-1$. As $N$ approaches $\infty$, the correlation coefficient between two SMFs should approach 0. We also perform the curve fitting on the simulation results and find that the correlation coefficients can be fitted as $-\frac{1}{N-1}$, which coincides to the analytical result obtained in Section \ref{TruncatedGaussian}.

\subsubsection{Truncated Multivariate Gaussian Model For Power Distribution}\label{TruncatedGaussian}
According to above simulation results, it is reasonable to assume that the ratios $\bm{a}=[a_1, a_2, \cdots, a_N]^\text{T}$ for the optical power satisfies a truncated multivariate Gaussian distribution \cite{Wilhelm2010} over the simplex defined in \eqref{Simplex}. The mathematical expectation of this truncated multivariate Gaussian distribution is $E[\bm{a}]=\bm{\mu}_{\bm{a}}=[\frac{1}{N}, \frac{1}{N}, \cdots, \frac{1}{N}]^{\text{T}}$, where $[\cdot]^{\text{T}}$ represents the transpose operator \footnote{Our analysis can be easily extended to the cases where different SMF ends have different mean values by replacing $\bm{\mu}_{\bm{a}}$ with the actual mean values.}. Here we derive the PDF of this truncated multivariate Gaussian distribution analytically.

We first remove the constraint $a_1+a_2+\cdots+a_N=1$, then the joint PDF of the truncated multivariate Gaussian distribution has the following form:
\begin{equation}
\begin{aligned}
f(\bm{a})&=\frac{1}{C_1} \exp \left\{-\frac{1} {2} [\bm{a}-\bm{\mu}_{\bm{a}}]^\text{T} {\bm{\Sigma}}_{\bm{a}} ^{-1} [\bm{a}-\bm{\mu}_{\bm{a}}] \right\},\\
&\quad \quad  0 \leq a_i \leq 1, \quad i=1, 2, \cdots, N,
\label{eq: fbmx}
\end{aligned}
\end{equation}
\noindent where $C_1=\int_V \exp \left\{-\frac{1} {2} [\bm{a}-\bm{\mu}_{\bm{a}}]^\text{T} {\bm{\Sigma}}_{\bm{a}} ^{-1} [\bm{a}-\bm{\mu}_{\bm{a}}] \right\} \mathrm{d} V$ is a constant number for normalization; $V$ is the domain defined as $V=\{0 \leq a_i \leq 1,\ i=1, 2, \cdots, N \}$; $\bm{\Sigma}_{\bm{a}}$ is the covariance matrix of $\bm{a}$. Because this work assumes non-mode-selective PL, it is reasonable to assume that $a_1, a_2, \cdots, a_{N}$ have the same Gaussian variance \footnote{Note that the Gaussian variance $var(a_i)$ here is not the actual variance of $a_i$. This is because the multivariate Gaussian distribution characterized is truncated by the definition domain $V$. Then the actual variance $var_{Actual}(a_i)$ is defined as $var_{Actual}(a_i)\triangleq \int_V (a_i-1/N)^2 f(\bm{a})\mathrm{d}V $, which is smaller than the Gaussian variance $var(a_i)$.} $var(a_i)=\sigma^2, \ i=1, 2, \cdots, N$; and the Gaussian covariances $cov(a_i, a_j)$ for any $a_i$ and $a_j$, when $i \neq j, \ i, j=1, 2, \cdots, N$, are the same. Then the $N \times N$ dimensional Gaussian covariance matrix $\bm{\Sigma}_{\bm{a}}$ can be written as
\begin{equation}
\begin{aligned}
\bm{\Sigma}_{\bm{a}} = \sigma ^2
\left[
\begin{array}{ccccc}
1 & \rho & \cdots & \rho \\
\rho & 1 & \cdots & \rho \\
\vdots & \vdots & \vdots & \vdots\\
\rho & \rho & \cdots & 1 \\
\end{array}
\right],
\label{eq: bmSigma}
\end{aligned}
\end{equation}
\noindent where $\rho=\frac{cov(a_i, a_j)}{\sigma^2}$ is the correlation coefficient between $a_i$ and $a_j$ when $i \neq j, \ i, j=1, 2, \cdots, N$. Then inverse matrix $\bm{\Sigma}_{\bm{a}}^{-1}$ in \eqref{eq: fbmx} can be obtained as
\begin{equation}
\begin{aligned}
\bm{\Sigma}_{\bm{a}} ^{-1}&= \frac{1}{[1+(N-1)\rho](1-\rho)\sigma^2}\\
&\times
\left[ \begin{array}{cccccc}
1+(N-2)\rho & -\rho & \cdots & -\rho\\
-\rho & 1+(N-2)\rho & \cdots & -\rho\\
\vdots & \vdots & \vdots & \vdots\\
-\rho & -\rho & \cdots & 1+(N-2)\rho
\end{array}
\right].
\label{eq: AbmSigmainv}
\end{aligned}
\end{equation}

However, when the constraint $a_1+a_2+\cdots+a_N=1$ is considered, the covariance matrix $\bm{\Sigma}_{\bm{a}}$ becomes a rank-deficient matrix and it has no inverse matrix. We first derive the correlation coefficient $\rho$. The constraint $a_1+a_2+\cdots+a_N=1$ can be rewritten as $[\bm{a}-\bm{\mu}_{\bm{a}}]^\text{T} \bm{1} =0$, where $\bm{1}=[1, 1, \cdots, 1]^\text{T}$ is an $N\times 1$ dimensional vector. Then we have \cite{zhang2019study}
\begin{equation}
\begin{aligned}
\label{MidEq}
E\left[[\bm{a}-\bm{\mu}_{\bm{a}}][\bm{a}-\bm{\mu}_{\bm{a}}]^\text{T}\bm{1}\right]&=\bm{\Sigma}_{\bm{a}}\bm{1}\\
&=\sigma^2(1+(N-1)\rho)\bm{1}\\
&=\bm{0},
\end{aligned}
\end{equation}
\noindent where $\bm{0}=[0, 0, \cdots, 0]^\text{T}$ is an $N\times 1$ dimensional zero vector. Therefore, the correlation coefficient $\rho$ can be obtained from \eqref{MidEq} as $\rho=-\frac{1}{N-1}$, which is the same as the correlation coefficient obtained from the simulation result in Fig. \ref{fig: Correlation_Coefficient}. This correlation coefficient is also consistent with the inverse matrix in \eqref{eq: AbmSigmainv} because the numerator of $\bm{\Sigma}_{\bm{a}}^{-1}$ becomes zero when $\rho=-\frac{1}{N-1}$, and thus the inverse matrix does not exist.

To obtain the explicit form of the joint PDF, we generalize a generalized inverse matrix of $\bm{\Sigma}_{\bm{a}}$, and let ${\rho \to -\frac{1}{N-1}}$ when the constraint $a_1+a_2+\cdots+a_N=1$ is considered. Then the joint PDF of $\bm{a}$ can be obtained by substituting \eqref{eq: AbmSigmainv} into \eqref{eq: fbmx} and letting ${\rho \to -\frac{1}{N-1}}$. After some algebra (see Appendix \ref{Appendix A}), the joint PDF can be obtained as
\begin{equation}
\begin{aligned}
f(\bm{a})&=\frac{1}{C_2}
\exp \left\{ - \frac{1}{2} [\bm{a^*}-\bm{\mu}_{\bm{a^*}}]^\text{T} \bm{\Sigma}_{\bm{a^*}}^{-1} [\bm{a^*}-\bm{\mu}_{\bm{a^*}}]\right\}\\
& \quad \times \delta \left({{a}_{1}}+{{a}_{2}}+\cdots+{{a}_{N}}-1 \right),\\
& \quad 0 \leq a_i \leq 1, \ i=1, 2,\cdots, N,
\label{eq: fbmxF}
\end{aligned}
\end{equation}
\noindent where
\begin{equation}\label{C_2}
\begin{aligned}
C_2&=\int_V \exp \left\{ - \frac{1}{2} [\bm{a^*}-\bm{\mu}_{\bm{a^*}}]^\text{T} \bm{\Sigma}_{\bm{a^*}}^{-1} [\bm{a^*}-\bm{\mu}_{\bm{a^*}}]\right\}\\
&\quad \times
\delta \left({{a}_{1}}+{{a}_{2}}+\cdots+{{a}_{N}}-1 \right) \mathrm{d}V
\end{aligned}
\end{equation}
\noindent is a constant normalization factor; $\bm{a^{*}}=[a_1, a_2, \cdots, a_{N-1}]^\text{T}$ is an $(N-1) \times 1$ dimensional vector; $\bm{\mu}_{\bm{a^*}}=[\frac{1}{N}, \frac{1}{N}, \cdots, \frac{1}{N}]^\text{T}$ is an $(N-1) \times 1$ dimensional vector; and the covariance matrix $\bm{\Sigma}_{\bm{a^{*}}}$ for $\bm{a^*}$ is the first $(N-1)\times(N-1)$ dimensional submatrix of $\bm{\Sigma}_{\bm{a}}$. Then the inverse of $\bm{\Sigma}_{\bm{a^{*}}}$ can be obtained as
\begin{equation}\label{Inverse_a_star}
\begin{aligned}
\bm{\Sigma}_{\bm{a^*}}^{-1}&= \frac{N-1}{N\sigma^2}
\left[ \begin{array}{cccccc}
2 & 1 & \cdots & 1\\
1 & 2 & \cdots & 1\\
\vdots & \vdots & \vdots & \vdots\\
1 & 1 & \cdots & 2
\end{array}
\right].
\end{aligned}
\end{equation}
\subsection{Two Extreme Cases}
\label{TwoCases}
Here we present two extreme cases of the truncated Gaussian distribution: the (joint) degenerate distribution and the (joint) uniform distribution, corresponding to the cases of minimum Gaussian variance $\sigma^2=0$ and maximum Gaussian variance \footnote{The actual variance of $a_i$ is $\frac{1}{12}$, because $a_i$ is uniformly distributed on $[0,1]$.} $\sigma^2=\infty$, respectively.

\subsubsection{Degenerate Distribution Case}
For a degenerate distribution, $a_i=\frac{1}{N}, i=1, 2, \cdots, N$ with probability one. Therefore, the joint PDF can be expressed as
\begin{equation}
\begin{aligned}
f(\bm{a})&=\Pi_{i=1}^{N}\delta(a_i-\frac{1}{N}).
\label{eq: degeneratePDF}
\end{aligned}
\end{equation}

\subsubsection{Uniform Distribution Case}
For an uniform distribution, $a_i, i=1, 2, \cdots, N$ is uniformly distributed on $[0, 1]$. The explicit joint PDF can be obtained by letting $\sigma^2\to\infty$ in \eqref{eq: fbmxF}, i.e.,
\begin{equation}
\begin{aligned}
f(\bm{a})&=\lim_{\sigma^2 \to \infty}\frac{\exp \left\{ - \frac{1}{2} [\bm{a^*}-\bm{\mu}_{\bm{a^*}}]^\text{T} \bm{\Sigma}_{\bm{a^*}}^{-1} [\bm{a^*}-\bm{\mu}_{\bm{a^*}}]\right\}\delta \left({{a}_{1}}+{{a}_{2}}+\cdots+{{a}_{N}}-1 \right)}{C_2}.
\label{eq: uniformPDF_1}
\end{aligned}
\end{equation}

From the expression of $\bm{\Sigma}_{\bm{a^*}}^{-1}$ in \eqref{Inverse_a_star}, we can find that the exponential term in \eqref{eq: uniformPDF_1} approaches one when $\sigma^2 \to \infty$. Similarly, for the denominator $C_2$, we have
\begin{equation}
\begin{aligned}
\lim_{\sigma^2 \to \infty}C_2&= \int_V \delta \left({{a}_{1}}+{{a}_{2}}+\cdots+{{a}_{N}}-1 \right) \mathrm{d}V\\
&=\int_{V_{s}} \mathrm{d}V_s\\
&= \frac{1}{(N-1)!},
\label{eq: C_2_uniform}
\end{aligned}
\end{equation}
\noindent where $V_s=\frac{1}{(N-1)!}$ is the volume of the standard simplex defined in \eqref{Simplex}. Substituting \eqref{eq: C_2_uniform} into \eqref{eq: uniformPDF_1}, we can obtain the joint PDF as
\begin{equation}
\begin{aligned}
f(\bm{a})&=(N-1)!\delta \left({{a}_{1}}+{{a}_{2}}+\cdots+{{a}_{N}}-1 \right).
\label{eq: uniformPDF_2}
\end{aligned}
\end{equation}

\section{Signal-to-Noise Ratio}
\label{section: SNR}
We assume that the shot noise is the dominated noise source in the coherent receiver, which is reasonable due to the presence of high intensity LO beams in balanced photodetectors. When the output signals of the SMF ends are combined using EGC \footnote{Here EGC method is used to combine the output signals of all SMF ends of the PL. Because there is only one receiving port and no diversity technique is introduced, the name ``EGC" should not be confused with the diversity combining technique EGC in wireless communications.}, the instantaneous SNR can be obtained as \cite{zhang2019study}
\begin{equation}
\gamma^{EGC}_{PL}=\frac{R\eta_S\left(\sum_{i=1}^{N}\sqrt{P_{S,i}}\right)^2}{NqB},
\end{equation}
\noindent where $R$ is the responsivity of the photodiode; $\eta_S$ is the mixing efficiency of SMF; $q$ is the electronic charge and $B$ is the noise equivalent bandwidth of the detector. Substituting ${{P}_{S, i}}$ into $\gamma_{PL}$, we can obtain
\begin{equation}\label{SNR_PL}
\gamma^{EGC}_{PL}=K\left(\sum_{i=1}^{N}\sqrt{a_i}\right)^2 I,
\end{equation}
\noindent where $K=\frac{R A \zeta_M\xi_{PL}\eta_s}{NqB}$.

Then the average SNR of coherent FSOC system using PL based receiver with EGC is
\begin{equation}
\begin{aligned}
\bar{\gamma}^{EGC}_{PL}&=E[{\gamma_{PL}}]\\
&=K E\left[\left(\sum_{i=1}^{N}{\sqrt{a_i}}\right)^{2}\right]E[I]\\
&=K \left[1+N(N-1)E \left[\sqrt{a_1 a_2}\right]\right],
\end{aligned}
\label{eq: gammaEGCave}
\end{equation}
\noindent where we have used the assumption $E[I]=1$ and the equality $E\left[\left(\sum_{i=1}^{N}{\sqrt{a_i}}\right)^{2}\right]=1+N(N-1)E \left[\sqrt{a_1 a_2}\right]$.

Similarly, the instantaneous and average SNR for SC can be respectively obtained as
\begin{equation}\label{SNR_SC}
\begin{aligned}
\gamma^{SC}_{PL}&=KN\max_{i}\{a_i\} I
\end{aligned}
\end{equation}
\noindent and
\begin{equation}\label{Average_SNR_SC}
\begin{aligned}
\bar{\gamma}^{SC}_{PL}&=KNE\left[\max_{i}\{a_i\}\right].
\end{aligned}
\end{equation}
The instantaneous and average SNR for MRC can be respectively obtained as
\begin{equation}\label{SNR_MRC}
\begin{aligned}
\gamma^{MRC}_{PL}&=KN\sum_{i=1}^{N}a_i I\\
&=KNI
\end{aligned}
\end{equation}
\noindent and
\begin{equation}
\begin{aligned}
\bar{\gamma}^{MRC}_{PL}&=KN.
\end{aligned}
\end{equation}

An important observation is that the instantaneous and average SNR for MRC are irrelevant to the power distribution of the PL due to the relation $\sum_{i=1}^{N}a_i=1$. Since the BER is determined by the instantaneous SNR, we can also conclude that the power distribution of the PL has no effect on the BER performance of the PL based receiver when MRC is used. This is an unique feature of the MRC for PL based receiver. However, the MRC requires the measurements of both the amplitude the phase of the signal in each branch, which is more complex compared with other combining techniques. In practical implementation, EGC and SC are two widely used combining techniques.

For the average SNR of EGC or SC, it is challenging to obtain an analytical expression for $E \left[\sqrt{a_1 a_2}\right]$ in \eqref{eq: gammaEGCave} or $E\left[\max_{i}\{a_i\}\right]$ in \eqref{Average_SNR_SC} when a general truncated multivariate Gaussian distribution is considered. However, it is still meaningful to consider the extreme cases defined in \ref{TwoCases}, because the degenerate case and the uniform case correspond to the smallest and the largest variance of the signal strength in each branch, respectively.

\subsection{Degenerate Distribution Case}
For the degenerate distribution, $a_i=\frac{1}{N}, (i=1, 2, \cdots, N)$ and we have $E\left[\left(\sum_{i=1}^{N}{\sqrt{a_i}}\right)^{2}\right]=N$ and $E\left[\max_{i}\{a_i\}\right]=\frac{1}{N}$. Then the average SNR for EGC and SC become
\begin{equation}
\label{AverageSNR_Degenerate}
\bar{\gamma}^{EGC}_{PL,Deg}=K N
\end{equation}
\noindent and
\begin{equation}
\label{AverageSNR_SC_Degenerate}
\bar{\gamma}^{SC}_{PL,Deg}=K.
\end{equation}
\noindent The average SNR of the EGC in degenerate distribution case equals the average SNR of the MRC. This is because all the branches have the same signal strength, then the EGC becomes an MRC.

\subsection{Uniform Distribution Case}
For the uniform distribution, $a_i, (i=1, 2, \cdots, N)$ is uniformed distributed in $[0,1]$ and we can obtain $E \left[\sqrt{a_1 a_2}\right]=\frac{\pi}{4 N}$ (see Appendix \ref{Appendix B}). Then the average SNR for EGC becomes
\begin{equation}
\label{AverageSNR_Uniform}
\bar{\gamma}^{EGC}_{PL,Uni}=K \frac{\pi N+4-\pi}{4}.
\end{equation}
However, it is still challenging to obtain an analytical form of average SNR of SC for uniform distribution case, except for the case with $N=2$. When $N=2$, we have $E\left[\max_{i}\{a_i\}\right]=\frac{3}{4}$. When $N>2$, we can use the Monte-Carlo method to numerically obtain $E\left[\max_{i}\{a_i\}\right]$ and then use the curve fitting method to approximate $E\left[\max_{i}\{a_i\}\right]$ as $\frac{4.45}{N+4.33}$. Then the average SNR for SC can be obtained as
\begin{equation}
\label{AverageSNR_SC_Uniform}
\bar{\gamma}^{SC}_{PL,Uni}\approx K \frac{4.45N}{N+4.33}.
\end{equation}

\subsection{General Distribution Case}
For a general truncated multivariate Gaussian distribution, i.e., $0<\sigma^2<\infty$, the average SNR is between the SNR of degenerate distribution and the SNR of uniform distribution. Now we consider the average SNR ratio of the degenerate distribution over the uniform distribution.

For the EGC, we have
\begin{equation}
\frac{\bar{\gamma}^{EGC}_{PL,Deg}}{\bar{\gamma}^{EGC}_{PL,Uni}}=\frac{4N}{\pi N+4-\pi},
\end{equation}
\noindent which is between $\frac{8}{4+\pi}\approx 1.12$ when $N=2$ and $\frac{4}{\pi} \approx 1.27$ when $N=\infty$. This implies that the influence of the power distribution of PL on the average SNR is relatively small when EGC method is used for signal combining.

For the SC, we have
\begin{equation}
\frac{\bar{\gamma}^{SC}_{PL,Deg}}{\bar{\gamma}^{SC}_{PL,Uni}}\approx \frac{N+4.33}{4.45 N},
\end{equation}
which is between $0.667$ when $N=2$ and $0.225$ when $N=\infty$. Since $\frac{\bar{\gamma}^{SC}_{PL,Deg}}{\bar{\gamma}^{SC}_{PL,Uni}}$ is always smaller than one, an interesting observation is that the SC prefers a large variance $\sigma^2$ than a small one. This is because the SC selects the largest $a_i$ as the output, then a larger variance of $a_i$ can have a larger possibility of obtaining a large $a_i$.

\subsection{Signal-to-Noise Ratios for SMF and MMF receivers}
For comparison, we also present the SNR of the SMF receiver and MMF receiver here. When shot noise is the dominated noise, the instantaneous SNR of the SMF receiver is
\begin{equation}
\gamma_{SMF}=\frac{\zeta_S \eta_S R A}{qB} I,
\end{equation}
\noindent where $\zeta_S$ is the coupling efficiency of SMF; and the average SNR of SMF is
\begin{equation}
\bar{\gamma}_{SMF}=E[\gamma_{SMF}]=\frac{\zeta_S \eta_S R A}{qB}.
\end{equation}

Similarly, the instantaneous SNR of the MMF receiver is
\begin{equation}
\gamma_{MMF}=\frac{\zeta_M \eta_M R A}{qB} I,
\end{equation}
\noindent where $\eta_M$ is the mixing efficiency of MMF mixer; and the average SNR of MMF is
\begin{equation}
\bar{\gamma}_{MMF}=E[\gamma_{MMF}]=\frac{\zeta_M \eta_M R A}{qB}.
\end{equation}

\section{Bit-Error Rate}
\label{section: BER Analysis}
\subsection{Integral Expression of BER}
The BER conditioned on received signal irradiance $I$ and power distribution $\bm{a}$ for an FSOC BPSK system \footnote{Although we only present the BER for BPSK scheme here, the BER and symbol error rate (SER) for other coherent modulation schemes can be easily found in a similar way.} using PL based receiver is given by \cite{song2012error}
\begin{equation}
P_{e,PL}(I,\bm{a})=Q(\sqrt{\gamma_{PL}}),
\end{equation}
\noindent where $Q(\cdot)$ is the Gaussian $Q$-function; and $\gamma_{PL}$ is the instantaneous SNR, which can be obtained in \eqref{SNR_PL}, \eqref{SNR_SC}, and \eqref{SNR_MRC} for EGC, SC, and MRC, respectively. Then the unconditional BER for PL based receiver can be obtained as the following integral form
\begin{equation}
P_{e,PL}=\int_{0}^{\infty}\int_V f(I) f(\bm{a}) Q(\sqrt{\gamma_{PL}})\mathrm{d}\bm{a}\mathrm{d}I,
\label{eq: Pe_EGC}
\end{equation}
\noindent where $f(I)$ is the PDF of the signal irradiance $I$ given in \eqref{eq: f}, and $f(\bm{a})$ is the joint PDF of power ratios $\bm{a}$ given in \eqref{eq: fbmxF}.

Similarly, the unconditional BERs for SMF receiver and MMF receiver are obtained as
\begin{equation}
P_{e,SMF}=\int_{0}^{\infty}f(I)Q(\sqrt{\gamma_{SMF}})\mathrm{d}I
\end{equation}
\noindent and
\begin{equation}
P_{e,MMF}=\int_{0}^{\infty}f(I)Q(\sqrt{\gamma_{MMF}})\mathrm{d}I,
\end{equation}
\noindent respectively.

\subsection{Analytical Lower Bound For BER}
\label{section: Series Solution}

Because $\left(\sum_{i=1}^{N}\sqrt{\hat{a}_i}\right)^2 \leq N$, where the equal sign is obtained when $a_i=\frac{1}{N}, i=1,2,\cdots,N$, we have $\gamma^{EGC}_{PL}\leq KNI=\gamma^{MRC}_{PL}$. Besides, noting that $\max_{i}\{a_i\}\leq 1$, we have $\gamma^{SC}_{PL}\leq KNI=\gamma^{MRC}_{PL}$. Therefore, the SNR of PL based receiver is bounded by the SNR of MRC \footnote{This coincides with the fact that MRC is the optimal combining regarding the SNR performance.}. Then we can obtain a lower bound for $P_{e,PL}$ as
\begin{equation}
\begin{aligned}
P_{e,PL}^{lower}&=\int_{0}^{\infty}\int_V f(I) f(\bm{a})Q\left(\sqrt{KN I}\right)\mathrm{d}\bm{a}\mathrm{d}I\\
&=\int_{0}^{\infty}f(I) Q(\sqrt{\bar{\gamma}^{MRC}_{PL}I})\mathrm{d}I,
\label{P_e_lowerbound}
\end{aligned}
\end{equation}
\noindent which is also the unconditional BER of PL based receiver using MRC.

Then we can obtain an analytical expression of the lower bound \eqref{P_e_lowerbound} by using a series expansion of the modified Bessel function of the second kind in \eqref{eq: f} as \cite{song2012error}
\begin{equation}
\begin{aligned}
K_{v}(x)=&\frac{\pi}{2 \sin (\pi v)}\sum\limits_{p=0}^{\infty}{\left[\frac{(x/2)^{2p-v}}{\Gamma(p-v+1)p!}-\frac{(x/2)^{2p+v}}{\Gamma(p+v+1)p!}\right]}, \\
&\quad \quad \quad v\notin Z, \left| x \right|<\infty
\label{eq: KvX}
\end{aligned}
\end{equation}
\noindent and an alternative expression of the $Q$-function \cite{park2010average}
\begin{equation}
Q(x)=\frac{1}{\pi}\int_{0}^{\pi/2} \exp \left(-\frac{x^2}{2 \sin^2 \theta}\right)\mathrm{d}\theta.
\label{Qfunc}
\end{equation}

Substituting \eqref{eq: f}, \eqref{eq: KvX}, and \eqref{Qfunc} into \eqref{P_e_lowerbound}, and after some algebra (see Appendix \ref{Appendix C}), we can obtain an analytical lower bound in series form as
\begin{equation}
\begin{aligned}
P_{e, PL}^{lower}&=\frac{\Lambda\left(\alpha, \beta\right)}{2}
\sum_{p=0}^{\infty}\left\{{a_p\left(\alpha, \beta\right)}\left(\frac{{\bar{\gamma}^{MRC}_{PL}}}{2}\right)^{-\left({p+\beta}\right)}
B\left(\frac{1}{2}, p+\beta+\frac{1}{2}\right)\right. \\
&\quad \quad \quad -\left.{a_p\left(\beta,\alpha\right)}
\left(\frac{{\bar{\gamma}^{MRC}_{PL}}}{2}\right)^{-\left({p+\alpha}\right)}
B\left(\frac{1}{2}, p+\alpha+\frac{1}{2}\right)\right\},
\label{eq: PeEGC2}
\end{aligned}
\end{equation}
\noindent where $B(x,y)=\int_0^1t^{x-1}(1-t)^{y-1}\mathrm{d}t$ is the Beta function, and
\begin{equation}\label{LambdaAnda_p}
\begin{aligned}
&\Lambda(\alpha, \beta)=\frac{1}{\Gamma{(\alpha)}\Gamma{(\beta)} \sin [(\alpha-\beta)\pi]};\\
&a_p(x,y)=\frac{(xy)^{p+y}{\Gamma\left({p+y}\right)}}{\Gamma(p-x+y+1)p!}.
\end{aligned}
\end{equation}

In addition, by replacing $\bar{\gamma}^{MRC}_{PL}$ in \eqref{eq: PeEGC2} with $\bar{\gamma}_{SMF}$ and $\bar{\gamma}_{MMF}$, we can obtain the unconditional BER for SMF receiver and MMF receiver, respectively.

\subsection{Truncation Error Analysis}
To implement the series form lower bound BER in \eqref{eq: PeEGC2}, we have to truncate the summation of infinite terms into a summation of finite terms. Therefore, it is necessary to analyze the truncation error. For simplicity, in the following we use $\bar{\gamma}$ to represent $\bar{\gamma}^{MRC}_{PL}$, $\bar{\gamma}_{SMF}$, and $\bar{\gamma}_{MMF}$. Substituting $B\left(x, y\right)=\frac{\Gamma\left(x\right)\Gamma\left(y\right)}{\Gamma\left(x+y\right)}$
\cite[ 8.384(1)]{gradshteyn2014table} and $\Gamma(x+1)=x\Gamma(x)$ \cite[ 8.331(1)]{gradshteyn2014table} into \eqref{eq: PeEGC2}, we obtain the error probability as
\begin{equation}
\begin{aligned}
P_{e}= \frac{\sqrt{\pi}}{2}\Lambda\left(\alpha, \beta\right)
\sum_{p=0}^{\infty}\frac{1}{p!}\left(\frac{2\alpha\beta}{\bar{\gamma}}\right)^p
\left\{G_p{\left(\alpha, \beta\right)}-G_p{\left(\beta, \alpha\right)}\right\},
\label{eq: PeEGC3}
\end{aligned}
\end{equation}
\noindent where $G_p{(x, y)}$ is defined as
\begin{equation}
G_p{(x,y)}=\frac{\Gamma(p+y+\frac{1}{2})}{(p+y)\Gamma(p-x+y+1)}\left(\frac{2xy}{\bar{\gamma}}\right)^y.
\end{equation}

Now we can estimate the truncation error caused by eliminating the infinite terms after the first $J$ terms in \eqref{eq: PeEGC3}. This truncation error can be defined as
\begin{equation}
\begin{aligned}
{\epsilon}_{J}=\frac{\sqrt{\pi}}{2}\Lambda\left(\alpha, \beta\right)
\sum_{p=J}^{\infty}\frac{1}{p!}\left(\frac{2\alpha\beta}{\bar{\gamma}}\right)^p
\left\{G_p{\left(\alpha, \beta\right)}-G_p{\left(\beta, \alpha\right)}\right\}.
\end{aligned}
\end{equation}

When $ p\to \infty$, we have $G_p{\left(\alpha, \beta\right)} \to 0$ and $G_p{\left(\beta, \alpha\right)} \to 0$. Then we can obtain an upper bound of the truncation error as
\begin{equation}
\begin{aligned}
{\epsilon}_{J}&<\frac{\sqrt{\pi}}{2}\Lambda\left(\alpha, \beta\right)
\sum_{p=J}^{\infty}\frac{1}{p!}\left(\frac{2\alpha\beta}{\bar{\gamma}}\right)^p
\max_{p\geq L}\left\{G_p{\left(\alpha, \beta\right)}-G_p{\left(\beta, \alpha\right)}\right\}\\
&\quad \quad <\frac{\sqrt{\pi}}{2}\frac{\Lambda\left(\alpha, \beta\right)}{J!}\left(\frac{2\alpha\beta}{\bar{\gamma}}\right)^J
\exp\left(\frac{2\alpha\beta}{\bar{\gamma}}\right)\\
&\quad \quad \quad \quad \times \max_{p\geq L}\left\{G_p{\left(\alpha, \beta\right)}-G_p{\left(\beta, \alpha\right)}\right\},
\end{aligned}
\end{equation}
\noindent where in the last inequality we have used the Lagrange form for the remainder term of Taylor series expansion for the exponential function. Note that when $J$ approaches $\infty$, the term $\frac{1}{J!}\left(\frac{2\alpha\beta}{\bar{\gamma}}\right)^J$ approaches zero. Therefore, truncation error ${\epsilon}_{J}$ diminishes to zero with increasing index $J$. Besides, we can also observe that ${\epsilon}_{J}$ diminishes rapidly with the average SNR $\bar{\gamma}$. This suggests that the series lower bound solution is highly accurate in the large SNR regimes. We can therefore perform an asymptotic BER analysis.

\subsection{Asymptotic Lower Bound For BER}
We now examine the lower bound BER behavior in the large SNR regimes. When $\bar{\gamma} \to \infty$, we have $G_p(\alpha,\beta) \to 0$ and $G_p(\beta,\alpha) \to 0$. From \eqref{eq: PeEGC3} we know that the first term $(p=0)$ of the series summation becomes the dominant term in the large SNR regimes. Therefore, the unconditional lower bound BER in high SNR regimes can be approximated by
\begin{equation}
\begin{aligned}
P_{e} \approx \frac{\sqrt{\pi}}{2}\Lambda\left(\alpha, \beta\right) \left[G_0{\left(\alpha, \beta\right)}-G_0{\left(\beta, \alpha\right)}\right].
\label{eq: PeEGCA}
\end{aligned}
\end{equation}

For typical turbulence conditions, we have $\alpha>\beta$. Then in high SNR regimes, we have
\begin{equation}
\begin{aligned}
\frac{G_0{\left(\beta, \alpha\right)}}{G_0{\left(\alpha, \beta\right)}}
=\frac{\beta \Gamma(\alpha+\frac{1}{2})\Gamma(-\alpha+\beta+1)}{\alpha \Gamma(\beta+\frac{1}{2})\Gamma(-\beta+\alpha+1)}
\left(\frac{2\alpha\beta}{\bar{\gamma}}\right)^{\alpha-\beta} \ll 1.
\end{aligned}
\end{equation}

Therefore, we can omit the second term in \eqref{eq: PeEGCA} and obtain
\begin{equation}
\begin{aligned}
P_{e} \approx H(\alpha, \beta) \left(\frac{1}{\bar{\gamma}}\right)^{\beta},
\label{eq: PeAsymError}
\end{aligned}
\end{equation}
\noindent where
\begin{equation}
H(\alpha, \beta)=\frac{\sqrt{\pi}}{2}\frac{\left({2\alpha\beta}\right)^{\beta}{\Gamma\left(\beta+\frac{1}{2}\right)}}{{\Gamma{(\alpha)}}{\Gamma\left(\beta+1\right)}
{\Gamma\left(-\alpha+\beta+1\right)} \sin [(\alpha-\beta)\pi]}.
\end{equation}

This indicates that the asymptotic lower bound BERs in high SNR regimes of the coherent optical communication system based on PL receiver, SMF receiver, and MMF receiver are decayed exponentially by the average SNR with an exponential decay constant $\beta$.

\section{Numerical Results}
\label{section: Simulation}

When the PL power distribution satisfies a general truncated multivariate Gaussian distribution and $N$ is large, we have to count on the stochastic numerical integration methods to calculate the normalization constant $C_2$, the average SNR and unconditional BER. However, the generation of random numbers satisfying truncated multivariate Gaussian distribution is not trivial. Here we use the Monte-Carlo integration (MCI) method (see Appendix \ref{Appendix D}) to calculate them. We set the number of SMF ends as $N=10$ in the following simulations. The turbulence parameters are $\left(\alpha =2.23, \ \beta =1.54\right)$ for moderate turbulence condition and $\left(\alpha =2.34, \ \beta =1.02\right)$ for strong turbulence condition \cite{al2001mathematical}. For the PL based receiver, we consider three different Gaussian variances $\sigma^2=0$, $\sigma^2=0.01$, and $\sigma^2=\infty$, which corresponds to the multivariate degenerate distribution, general truncated multivariate Gaussian distribution, and multivariate uniform distribution, respectively.

\begin{figure}
\begin{center}
\includegraphics [width=0.6\textwidth, draft=false] {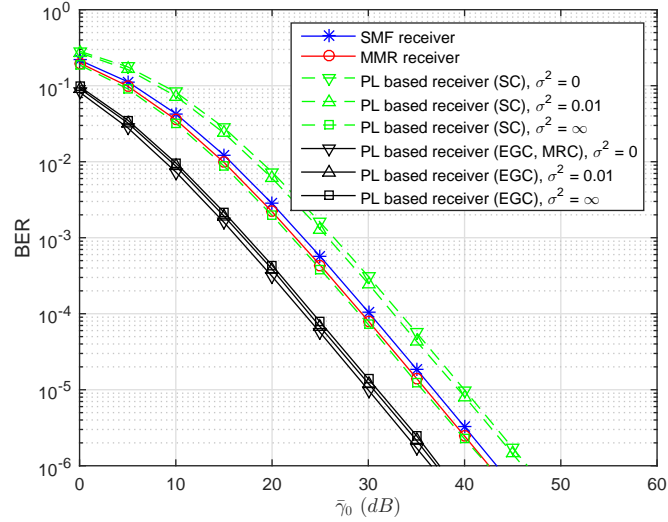}
\caption{The BER comparison between PL based receiver, SMF receiver, and MMF receiver for coherent FSOC system ($\xi_{PL}=0.8$, $\frac{\eta_S}{\eta_M}=5$, $\frac{\zeta_M}{\zeta_S}=6$)}
\label{BERComparison_different combining}
\end{center}
\end{figure}

We first present the BERs of SMF receiver, MMF receiver, and PL based receiver using different combining techniques under moderate turbulence, shown in Fig. \ref{BERComparison_different combining}. We take the SMF receiver as the reference and the horizontal axis is the average SNR of the SMF receiver, i.e., $\bar{\gamma}_0=\bar{\gamma}_{SMF}$. As we have demonstrated in Sections \ref{section: SNR} and \ref{section: BER Analysis}, the BER for MRC is irrelevant with the power distribution of the PL and it equals to the BER for EGC with degenerate distribution. From Fig. \ref{BERComparison_different combining}, we can see that the BER performance for EGC is much better than that for SC. Besides, the BERs for EGC under different PL power distributions are close to each other. The ratio of BER of $\sigma^2=\infty$ over BER when $\sigma^2=0$ is around $1.3$. This indicates that the PL power distribution has limited influence on the BER performance of the PL based receiver when EGC is used. However, the BER for SC of the uniform distribution ($\sigma^2=\infty$) is much lower than that of the degenerate distribution ($\sigma^2=0$). The ratio of BER when $\sigma^2=0$ over BER when $\sigma^2=\infty$ is around $4.8$. This indicates that the PL power distribution can greatly affect the BER performance of the PL based receiver when SC is used.

Because the BERs for EGC under different PL power distributions are close to the BER for MRC and they are much better than the BER for SC under different PL power distributions, next we will focus on the performance comparison between the PL based receiver with EGC, the SMF receiver, and the MMF receiver.

\begin{figure}
\begin{center}
\includegraphics [width=0.6\textwidth, draft=false] {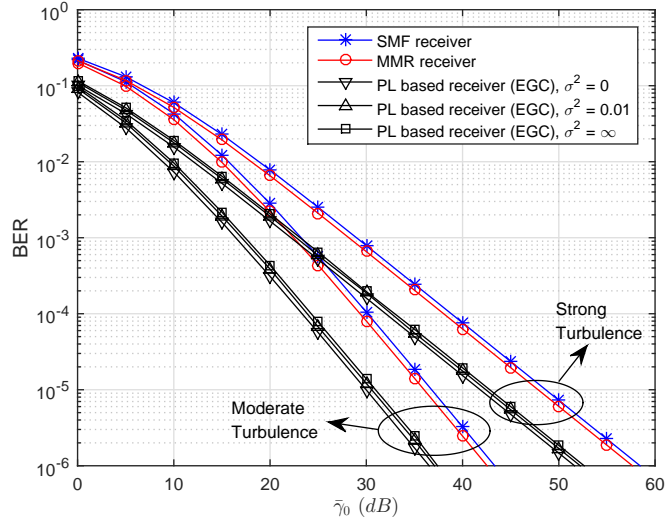}
\caption{The BER comparison between PL based receiver, SMF receiver, and MMF receiver for coherent FSOC system ($\xi_{PL}=0.8$, $\frac{\eta_S}{\eta_M}=5$, $\frac{\zeta_M}{\zeta_S}=6$)}
\label{BERComparison}
\end{center}
\end{figure}

The BER comparison between the PL based receiver with EGC, the SMF receiver, and the MMF receiver for coherent FSOC systems under the moderate and strong turbulence conditions is shown in Fig. \ref{BERComparison}. From Fig. \ref{BERComparison}, we can find that, when the BER is $10^{-6}$, the ${\bar{\gamma}_{0}}$ for the PL based receiver with EGC, the MMF receiver, and the SMF under moderate turbulence are about $37$ dB, $42.5 $ dB and $43$ dB, respectively; and under strong turbulence are about $53$ dB, $57.5$ dB and $58$ dB, respectively. This suggests that SMF receiver and MMF receiver require an additional $6$ dB and $5.5$ dB SNR to achieve the same BER as the PL based receiver with EGC under moderate turbulence; and require an additional $5$ dB and $4.5$ dB SNR to achieve the same BER as the PL based receiver with EGC under strong turbulence.

Then we present the average SNR gains of the PL based receiver with EGC over the SMF receiver
$\bar{\gamma}_{PL}/\bar{\gamma}_{SMF}=\frac{\xi_{PL}}{N}\frac{\zeta_M}{\zeta_S}E\left[\left(\sum\limits_{i=1}^{N}\sqrt{a_i}\right )^{2}\right]$, and over the MMF receiver $\bar{\gamma}_{PL}/\bar{\gamma}_{MMF} =\frac{\xi_{PL}}{N}\frac{\eta_S}{\eta_M}E\left[\left (\sum\limits_{i=1}^{N}\sqrt{a_i}\right)^{2}\right]$ under various imperfect device parameters, including the coupling efficiency, the mixing efficiency, and the PL loss. We analyze the value of the coupling efficiencies of MMF, few-mode fiber and SMF in the literature \cite{winzer1998fiber,dikmelik2005fiber,toyoshima2006maximum,grein2014multimode,poliak2016fiber,arisa2017coupling,hu2018fiber,ozdur2015photonic,zheng2016free}, and set the coupling efficiency gain of MMF over SMF as $\frac{\zeta_M}{\zeta_S} \in [0,20]$. We analyze the value of the mixing efficiency of SMF and MMF in the literature \cite{mark1992comparison,duncan1993performance,jacob1995heterodyne,Marshalek1996,leeb1998aperture,ren2012heterodyne}, and set the mixing efficiency gain of SMF over MMF as $\frac {{\eta}_{S}} {{\eta}_{M}} \in [4,8]$. The range of the PL loss is set as $\xi_{PL} \in [0,1]$.

\begin{figure}
\begin{center}
\includegraphics [width=0.6\textwidth, draft=false] {{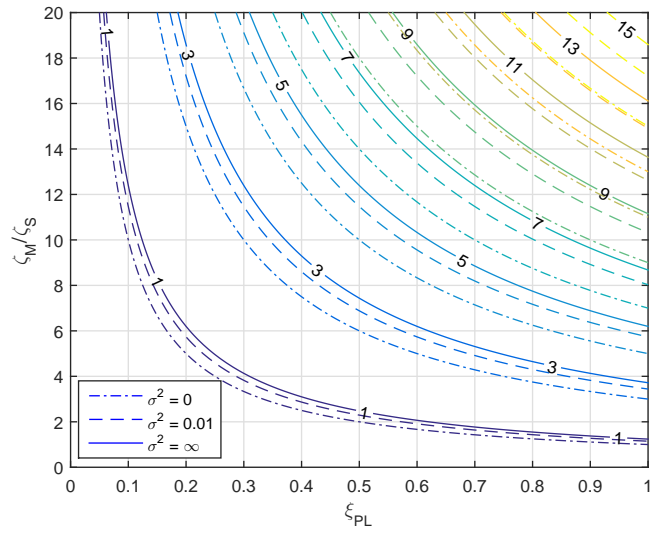}}
\caption{The average SNR gain of PL based receiver over SMF receiver for coherent FSOC system}
\label{fig: loss_coupling_S}
\end{center}
\end{figure}

\begin{figure}
\begin{center}
\includegraphics [width=0.6\textwidth, draft=false] {{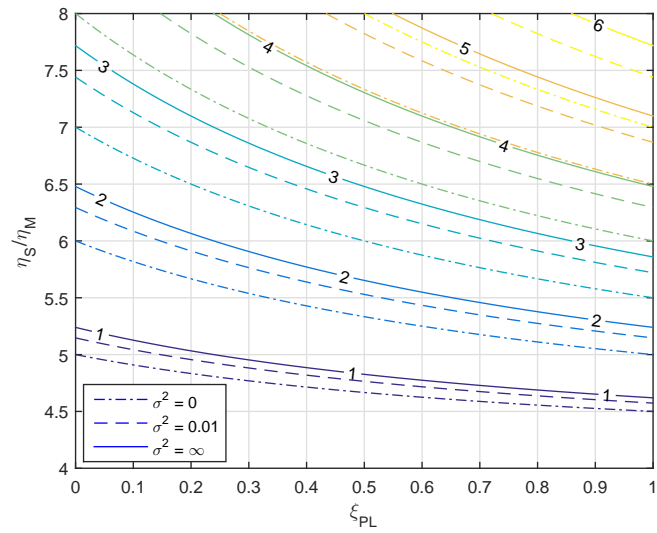}}
\caption{The average SNR gain of PL based receiver over MMF receiver for coherent FSOC system}
\label{fig: loss_coupling_M}
\end{center}
\end{figure}

Figures \ref{fig: loss_coupling_S} and \ref{fig: loss_coupling_M} show the obtained average SNR gain $\bar{\gamma}^{EGC}_{PL}/\bar{\gamma}_{SMF}$ and $\bar{\gamma}^{EGC}_{PL}/\bar{\gamma}_{MMF}$, respectively. The scope of application of the PL based receiver for FSOC systems can be obtained from Figs. \ref{fig: loss_coupling_S} and \ref{fig: loss_coupling_M}: when $\bar{\gamma}^{EGC}_{PL}/\bar{\gamma}_{SMF}>1$, we can choose to use the PL based receiver instead of SMF receiver for coherent FSOC systems; when $\bar{\gamma}^{EGC}_{PL}/\bar{\gamma}_{MMF}>1$, we can choose to use the PL based receiver instead of MMF receiver for coherent FSOC systems. Besides, from Figs. \ref{fig: loss_coupling_S} and \ref{fig: loss_coupling_M}, we can observe that the difference of the average SNR gain among three PL power distributions increases as average SNR gain increases. This indicates that the influence of the PL power distribution on the average SNR gain becomes significant in high SNR gain.

\begin{figure}
\begin{center}
\includegraphics [width=0.6\textwidth, draft=false] {{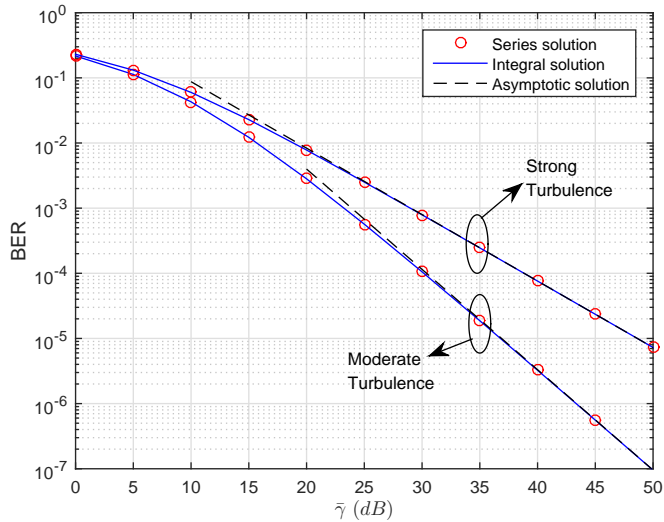}}
\caption{Comparison between integral solution, series solution, and asymptotic solution of coherent FSOC system using PL based receiver.}
\label{fig: BERaym}
\end{center}
\end{figure}

At last, we present the integral solution, series lower bound solution and the asymptotic solution of the unconditional lower bound BER in Fig. \ref{fig: BERaym}. The series lower bound solution is calculated by \eqref{eq: PeEGC2} with $J=30$. We can see that the series lower bound solution is consistent with the integral solution, and the asymptotic lower bound BER approaches the exact BER curve in high SNR regimes ($\bar{\gamma} >30$ dB).

\section{Conclusion}
\label{section: Conclusion}
This paper proposed a truncated multivariate Gaussian distribution over a simplex for the power distribution at SMF ends of the PL. The SNR and BER for PL based receiver are analyzed using different combining techniques, including SC, EGC, and MRC; and they are compared with the SMF and MMF receivers for FSOC systems. We demonstrated that the power distribution of the PL has no effect on the SNR and BER performance of PL based receiver when MRC is used. Simulation results showed that the power distribution of the PL has limited influence on the BER performance of PL based receiver when EGC is used; and it can greatly affect the BER performance of the PL based receiver when SC is used. Besides, we quantified the SNR gains of the PL based receiver using EGC over the SMF and MMF receivers under different imperfect devices parameters; and provided the scope of application of the considered communication system. These findings can provide some useful guidelines for the design of PL based receiver for FSOC systems.

We have to remark that the PL with $N$ SMF ends requires $N$ balanced photodetectors to detect the received beams. Therefore, the cost and the complexity of the PL based receiver are higher than those of the SMF receiver. To reduce the number of balanced photodetectors and lower the complexity of the receiver, in the future work we will combine hybrid combining techniques, e.g., the hybrid-selection/equal-gain combining \cite{ma2007unified}, with PL based receiver.

\section*{Acknowledgement}
Bo Zhang acknowledges the support from the UCAS Joint PhD Training Program that allows her to conduct this research while she was visiting The University of British Columbia, Canada.

\bibliographystyle{IEEEtran}
\bibliography{ref}

\appendices
\section{Derivation of $f(\bm{a})$}\label{Appendix A}
Substituting \eqref{eq: AbmSigmainv} into \eqref{eq: fbmx} and letting ${\rho \to -\frac{1}{N-1}}$, we can obtain
\begin{equation}
\begin{aligned}
f(\bm{a})=& \lim_{\rho \to -\frac{1}{N-1}} \frac{1}{C_1}
\exp \left \{ -\frac{1}{2} \frac{[1+(N-2)\rho]\sum\limits_{i=1}^{N}x_i^2
-2\rho \sum\limits_{i=1}^{N-1} \sum\limits_{j=i+1}^{N} x_i x_j}
{[1+(N-1)\rho](1-\rho)\sigma^2}\right \}\\
=&\underbrace {\lim_{\rho \to -\frac 1 {N-1}}\frac{1}{C_1}
\exp \left \{-\frac{1}{2}\frac{[1+(N-3)\rho]\sum\limits_{i=1}^{N-1}x_i^2
-{2\rho}\sum\limits_{i=1}^{N-1}\sum\limits_{j=i+1}^{N-2}x_i x_j}
{(1-\rho)[{1+(N-2)\rho}]\sigma^2}\right \}}_{H_1} \\
&\quad \quad \quad \times
\underbrace {\lim_{\rho \to -\frac{1}{N-1}}
\exp \left \{-\frac{1}{2} \frac{\left[x_N-\frac {\rho}{1+(N-2)\rho}\sum\limits_{i=1}^{N-1}x_i\right]^2}{\varepsilon^2}
\right \}}_{H_2},
\label{eq: Aexpa1}
\end{aligned}
\end{equation}
\noindent where $x_i=a_i-\frac{1}{N}$ and $\varepsilon=\sigma\sqrt{\frac{[1+(N-1)\rho](1-\rho)}{[1+(N-2)\rho]}}$.

When ${\rho \to -\frac{1}{N-1}}$, we have $\varepsilon \to 0 $. Because the limit of the Gaussian distribution can be expressed as the Dirac delta function, then we can simplify $H_2$ in \eqref{eq: Aexpa1} as
\begin{equation}
\begin{aligned}
H_2&=\lim_{\rho \to -\frac 1 {N-1}} {\sqrt{2\pi \varepsilon^{2}}}
\times
\delta \left[x_N-\frac {\rho}{1+(N-2)\rho}\sum\limits_{i=1}^{N-1}x_i\right]\\
&=\lim_{\rho \to -\frac 1 {N-1}}{\sqrt{2\pi \varepsilon^{2}}}
\times
\delta\left({{a}_{1}}+{{a}_{2}}+\cdots+{{a}_{N}}-1 \right),
\label{eq: xing}
\end{aligned}
\end{equation}
where $\delta(\cdot)$ is the Dirac delta function. Using the expression in \eqref{eq: AbmSigmainv}, we can simplify $H_1$ in \eqref{eq: Aexpa1} as
\begin{equation}
\begin{aligned}
H_1 =  \frac{1}{C_1}\exp \left\{ - \frac{1}{2} [\bm{a^*}-\bm{\mu}_{\bm{a^*}}]^\text{T} \bm{\Sigma}_{\bm{a^*}}^{-1} [\bm{a^*}-\bm{\mu}_{\bm{a^*}}]\right\}.
\label{eq: diamondsuit}
\end{aligned}
\end{equation}

Substituting \eqref{eq: xing} and \eqref{eq: diamondsuit} into \eqref{eq: Aexpa1}, we can obtain
\begin{equation}
\begin{aligned}
f(\bm{a})&= \lim_{\rho \to -\frac 1 {N-1}} \frac{\sqrt{2\pi \varepsilon^2}}{C_1}
\exp \left\{ - \frac{1}{2} [\bm{a^*}-\bm{\mu}_{\bm{a^*}}]^\text{T} \bm{\Sigma}_{\bm{a^*}}^{-1} [\bm{a^*}-\bm{\mu}_{\bm{a^*}}]\right\}\\
& \quad \quad \quad \quad \times \delta \left({{a}_{1}}+{{a}_{2}}+\cdots+{{a}_{N}}-1 \right).
\label{eq: Aexpa2}
\end{aligned}
\end{equation}

We can find that there exists the same factor ${\sqrt{2\pi \varepsilon^{2}}}$ in the numerator and denominator $C_1$. Finally, by eliminating the term ${\sqrt{2\pi \varepsilon^{2}}}$ in both numerator and denominator, the joint PDF in \eqref{eq: fbmxF} can be obtained.

\section{Derivation of $E \left[\sqrt{a_1 a_2}\right]$ For Uniform Distribution}\label{Appendix B}
The mathematical expectation of $\sqrt{a_1 a_2}$ for a joint PDF $f(\bm{a})$ in \eqref{eq: uniformPDF_2} is defined as
\begin{equation}
\begin{aligned}
E \left[\sqrt{a_1 a_2}\right]&=(N-1)!\int_0^1\sqrt{a_1}\int_0^{1-a_1}\sqrt{a_2}\int_0^{1-a_1-a_2}\cdots\int_0^{1-a_1-\cdots-a_{N-1}}\\
&\quad \quad \quad \times \delta \left({{a}_{1}}+{{a}_{2}}+\cdots+{{a}_{N}}-1 \right)\mathrm{d}a_1\mathrm{d}a_2\cdots\mathrm{d}a_N\\
&=(N-1)!\int_0^1\sqrt{a_1}\int_0^{1-a_1}\sqrt{a_2}\int_0^{1-a_1-a_2}\cdots\int_0^{1-a_1-\cdots-a_{N-2}}\mathrm{d}a_1\mathrm{d}a_2\cdots\mathrm{d}a_{N-1}.
\end{aligned}
\end{equation}
By integrating $a_{N-1}$ out, we can obtain
\begin{equation}
\begin{aligned}
E \left[\sqrt{a_1 a_2}\right]&=(N-1)!\int_0^1\sqrt{a_1}\int_0^{1-a_1}\sqrt{a_2}\int_0^{1-a_1-a_2}\cdots\int_0^{1-a_1-\cdots-a_{N-3}}\\
&\quad \quad \quad \times \frac{1}{1!}(1-a_1-a_2-\cdots-a_{N-2})\mathrm{d}a_1\mathrm{d}a_2\cdots\mathrm{d}a_{N-2}.
\end{aligned}
\end{equation}
By integrating $a_{N-2}$ out, we can obtain
\begin{equation}
\begin{aligned}
E \left[\sqrt{a_1 a_2}\right]&=(N-1)!\int_0^1\sqrt{a_1}\int_0^{1-a_1}\sqrt{a_2}\int_0^{1-a_1-a_2}\cdots\int_0^{1-a_1-\cdots-a_{N-4}}\\
&\quad \quad \quad \times \frac{1}{2!}(1-a_1-a_2-\cdots-a_{N-3})^2\mathrm{d}a_1\mathrm{d}a_2\cdots\mathrm{d}a_{N-3}.
\end{aligned}
\end{equation}
Similarly, by successively integrating $a_{N-3}, a_{N-4}, \cdots, a_3$ out, we can obtain
\begin{equation}
\begin{aligned}
E \left[\sqrt{a_1 a_2}\right]&=(N-1)!\int_0^1\sqrt{a_1}\int_0^{1-a_1}\sqrt{a_2}\frac{1}{(N-3)!}(1-a_1-a_2)^{N-3}\mathrm{d}a_1\mathrm{d}a_2\\
&=(N-1)(N-2)\int_0^1\sqrt{a_1}\int_0^{1-a_1}\sqrt{a_2}(1-a_1-a_2)^{N-3}\mathrm{d}a_1\mathrm{d}a_2.
\end{aligned}
\end{equation}

Using the relation of Beta function $\int_a^b(t-a)^{x-1}(b-t)^{y-1}\mathrm{d}t=(b-1)^{x+y-1}B(x,y)$ \cite[3.196(3)]{gradshteyn2014table}, we can obtain
\begin{equation}
\begin{aligned}
E \left[\sqrt{a_1 a_2}\right]&=(N-1)(N-2)B\left(\frac{3}{2},N-\frac{1}{2}\right)B\left(\frac{3}{2},N-2\right).
\end{aligned}
\end{equation}
Using the equalities $B(x,y)=\frac{\Gamma(x)\Gamma(y)}{\Gamma(x+y)}$ \cite[8.384(1)]{gradshteyn2014table}, $\Gamma(\frac{3}{2})=\frac{\pi}{4}$, and $\Gamma(m)=(m-1)!$, we can obtain
\begin{equation}
\begin{aligned}
E \left[\sqrt{a_1 a_2}\right]
&=\frac{\pi}{4N}.
\end{aligned}
\end{equation}

\section{Derivation of The Analytical Expression For $P_{e,PL}^{lower}$}\label{Appendix C}
Substituting \eqref{eq: f}, \eqref{eq: KvX}, and \eqref{Qfunc} into \eqref{P_e_lowerbound}, we can obtain
\begin{equation}
\begin{aligned}
P_{e,PL}^{lower}&=\Lambda(\alpha, \beta)\int_{0}^{\pi/2}\!\!\int_{0}^{\infty}
\left\{
\sum_{p=0}^{\infty}\left[\frac{a_p(\alpha, \beta){I}^{p+\beta-1}}{\Gamma\left({p+\beta}\right)}
\exp \left(-\frac{{\bar{\gamma}^{MRC}_{PL}}{I}}{2 \sin ^2{\theta}}\right)\right]\right. \\
&\quad \quad \quad \left.-\sum_{p=0}^{\infty}\left[\frac{a_p(\beta, \alpha){I}^{p+\alpha-1}}{\Gamma\left({p+\alpha}\right)}
\exp \left(-\frac{{\bar{\gamma}^{MRC}_{PL}} {I}}{2 \sin ^2{\theta}}\right)\right]
\right\}
\mathrm{d}{I}
\mathrm{d}\theta,
\label{eq: P}
\end{aligned}
\end{equation}
\noindent where $\Lambda(\alpha, \beta)$ and $a_p(x,y)$ are defined in \eqref{LambdaAnda_p}.

Using $\int_{0}^{\infty} {x^m \exp \left(-\beta x^n \right)}\mathrm{d}x = \frac{\Gamma(\frac{m+1}{n})}{n\beta^{\frac{m+1}{n}}}$ \cite[3.326(2)]{gradshteyn2014table}, we can obtain
\begin{equation}
\begin{aligned}
P_{e, PL, Deg}&=\Lambda(\alpha, \beta)\sum_{p=0}^{\infty}
\int_{0}^{\pi/2}\left\{{a_p\left(\alpha, \beta\right)}\left(\frac{{\bar{\gamma}^{MRC}_{PL}}}{2}\right)^{-\left({p+\beta}\right)}
\sin ^{2p+2\beta}{\theta}\right. \\
&\quad \quad \quad -\left.{a_p\left(\beta, \alpha\right)}\left(\frac{{\bar{\gamma}^{MRC}_{PL}}}{2}\right)^{-\left({p+\alpha}\right)}
\sin ^{2p+2\alpha}{\theta} \right\}\mathrm{d}\theta.
\label{PEG1}
\end{aligned}
\end{equation}

Using the Beta function $B\left(x, y\right)=2\int_0^{\pi/2} \sin ^{2x-1}\psi \cos ^{2y-1}\psi \mathrm{d}\psi$ \cite[8.380(2)]{gradshteyn2014table} and $B\left(x, y\right)=B\left(y, x\right)$ \cite[8.384(1)]{gradshteyn2014table} into (\ref{PEG1}), we can obtain the series solution to the unconditional BER as \eqref{eq: PeEGC2}.

\section{MCI Method For Calculating $C_2$, $\bar{\gamma}_{PL}$, $P_{out,PL}$, and $P_{e,PL}$}\label{Appendix D}
In an MCI method, to obtain the integral result of $\int_{\bm{x}} g(\bm{x}) \mathrm{d}\bm{x}$, we first choose a PDF $f(\bm{x})$, which is referred as the sampling function, and rewrite the integral as $\int_{\bm{x}} f(\bm{x}) \frac{g(\bm{x})}{f(\bm{x})}\mathrm{d}\bm{x}$. Then the integral can be viewed as the mathematical expectation of the objective function $O(\bm{x})\triangleq \frac{g(\bm{x})}{f(\bm{x})}$ when $\bm{x}$ subjects to a PDF $f(\bm{x})$, i.e., $\int_{\bm{x}} f(\bm{x}) O(\bm{x}) \mathrm{d}\bm{x}=E[\frac{g(\bm{x})}{f(\bm{x})}]$. Therefore, we can generate $M$ samples $\{\bm{x}_1, \bm{x}_2, \cdots, \bm{x}_M\}$ of $\bm{x}$ according to the PDF $f(\bm{x})$ and use the average value of the objective function $\frac{1}{M}\sum_{m=1}^M O(\bm{x}_m)$ to estimate the mathematical expectation \cite{yuan2019monte}.

For example, to obtain the normalization constant $C_2$, we can first rewrite $C_2$ as
\begin{equation}
C_2=\int_{-\infty}^\infty \cdots \int_{-\infty}^\infty f_{MG}(\bm{a}) C_3 I_n(\bm{a}) \mathrm{d}a_1\cdots \mathrm{d}a_N,
\end{equation}
\noindent where $C_3=[2\pi \sigma^2 N/(N-1)]^{\frac{N-1}{2}}/\sqrt{N}$ \cite{zhang2019study}; $f_{MG}(\bm{a})$ is the PDF of multivariate Gaussian variables $\bm{a}=[a_1, a_2, \cdots, a_N]^\text{T}$ satisfying $a_1+a_2+\cdots +a_N=1$, and $f_{MG}(\bm{a})$ is given by \cite{zhang2019study}
\begin{equation}
f_{MG}(\bm{a})=\frac{1}{C_3}\exp \left\{ - \frac{1}{2} [\bm{a^*}-\bm{\mu}_{\bm{a^*}}]^\text{T} \bm{\Sigma}_{\bm{a^*}}^{-1} [\bm{a^*}-\bm{\mu}_{\bm{a^*}}]\right\}\times \delta \left({{a}_{1}}+{{a}_{2}}+\cdots+{{a}_{N}}-1 \right)
\end{equation}
\noindent and $I_n(\bm{a})$ is an indicator function defined as
\begin{equation}
I_n(\bm{a})=\begin{cases}
1,&\text{$\bm{a} \in V$}\\
0,&\text{$\bm{a} \notin V$}.
\end{cases}
\end{equation}

Then we can choose $f_{MG}(\bm{a})$ as the sampling function and the objective function becomes $O(\bm{a})=C_3 I_n(\bm{a})$. The generation of random numbers $\{a_1,a_2,\cdots,a_{N}\}$ satisfying PDF $f_{MG}(\bm{a})$ can be achieved by two steps: first generate $\{a_1,a_2,\cdots,a_{N-1}\}$ according to the $N-1$ dimensional multivariate Gaussian PDF
\begin{equation}
f_{MG}(\bm{a^*})=\frac{1}{C_3}\exp \left\{ - \frac{1}{2} [\bm{a^*}-\bm{\mu}_{\bm{a^*}}]^\text{T} \bm{\Sigma}_{\bm{a^*}}^{-1} [\bm{a^*}-\bm{\mu}_{\bm{a^*}}]\right\};
\end{equation}
\noindent then $a_N$ is obtained as $a_N=1-a_1-a_2-\cdots-a_{N-1}$.

The average SNR and the unconditional BER can be calculated in a similar method. In addition, to calculate the average SNR and the unconditional BER for uniform distribution case, the key is to generate the random numbers $\{a_1, a_2, \cdots, a_N\}$ satisfying multivariate uniform distribution over the standard simplex. This can be achieved by the Algorithm 2 given in \cite{onn2011generating}.

\balance
\end{document}